\input{psfig}
\documentclass[]{aa}
\usepackage{graphicx}
\usepackage{deluxetable}
\usepackage{aalongtable}
\begin{document}
%\thesaurus{08(08.01.2, 08.09.2, 08.12.1, 08.13.1, 08.19.6)}
\title{RACE-OC Project:\\ Rotation and variability in the open cluster M\,11 (NGC\,6705)\thanks{Figs.\,\ref{a1}-\ref{n8} are only available in electronic form
at the CDS via anonymous ftp to cdsarc.u-strasbg.fr (130.79.128.5)
or via http://cdsweb.u-strasbg.fr/cgi-bin/qcat?J/A+A/}}
\author{S.\,Messina\inst{1}
\and          P.\, Parihar\inst{2}
\and          J.\,-R.\,Koo\inst{3}
\and          S.\,-L.\,Kim\inst{3}
\and          S.\,-C.\,Rey\inst{4}
\and          C.\,-U.\,Lee\inst{3}
}
\offprints{Sergio Messina}
\institute{INAF-Catania Astrophysical Observatory, via S.\,Sofia 78, I-95123 Catania, Italy  \\
\email{sergio.messina@oact.inaf.it}
\and   
Indian Institute of Astrophysics, Block II, Koramangala, Bangalore India, 560034 \\
\email{psp@iiap.res.in}
\and
Korea Astronomy and Space Science Institute,  Daejeon, Korea\\
\email{koojr@kasi.re.kr; slkim@kasi.re.kr}
\and
Department of Astronomy and Space Science, Chungnam National University, Daejeon, Korea
\\}
%\and
%Department of Physics and Astronomy, Johns Hopkins University,
%Baltimore, MD 21218, USA\\

\date{}
\titlerunning{Rotation and variability in M\,11}
\authorrunning{S.\,Messina et al.}
\abstract {Rotation  and  magnetic   activity  are  intimately  linked  in
main-sequence stars  of G  or later spectral  types. The  presence and
level  of magnetic  activity depend  on stellar  rotation,  and rotation
itself is strongly influenced by strength and topology of the magnetic
fields. Open clusters represent especially useful targets to investigate the rotation/activity/age connection. 
Over the time stellar activity and  rotation evolve,  providing us with
a promising diagnostic  tool to determine
age of the field stars.}
 { The open cluster \object{M11} has been studied as a part of the RACE-OC project ({\bf R}otation and {\bf AC}tivity {\bf E}volution in {\bf O}pen {\bf C}lusters), which is aimed at exploring the evolution 
of rotation and magnetic activity in the late-type members of open clusters with different ages.} {Photometric observations of the open cluster M11 were carried out in
June 2004  using LOAO 1m telescope. The  rotation periods of the
cluster members are determined by  Fourier analysis of 
photometric data time series.   We further investigated the  relations between the
surface  activity, characterized  by  the light  curve amplitude,  and
rotation.} {We have discovered a total of 75 periodic variables in the M11 FoV, of which 38 are candidate cluster members. Specifically, among cluster members  we discovered 6 early-type, 2 eclipsing binaries and 30 bona-fide single periodic late-type variables.
Considering the rotation periods of 16 G-type members of the almost coeval 200-Myr M34 cluster, we could determine the rotation period distribution
 from a more numerous sample of  46 single G stars at an age of about 200-230 Myr  and determine a median rotation period P=4.8d.} {A comparison with the younger
M35  cluster ($\sim$150 Myr) and with the older  \object{M37} cluster  ($\sim$550 Myr) shows that G stars rotate slower than younger M35 stars and faster than older M37 stars.
The measured  variation of the median rotation period is consistent with the scenario of rotational braking  of main-sequence spotted stars as they age. 
Finally, G-type \object{M11} members have a  level of photospheric magnetic activity, as measured by light curve amplitude,  comparable to that observed in the in younger 110-Myr 
Pleiades  stars of similar mass and rotation. }
\maketitle
\rm
\section{Introduction}
RACE-OC, which stands for {\bf R}otation and {\bf AC}tivity {\bf E}volution in {\bf O}pen {\bf C}lusters, is a long-term 
project with the aim to study stellar rotation, magnetic activity, and their evolution in the late-type members of open
clusters (Messina \cite{Messina07}; Messina et al. \cite{Messina08a}).
The project's targets are open clusters which, differently than field stars, provide us with a sample of stars spanning a range of masses with same age, initial chemical composition, 
environmental conditions, and interstellar reddening.
Such stellar samples allow us to accurately investigate the rotation/activity/age relashionships and their mass dependence.\\
\indent
Indeed, rotation is one  basic property of late-type stars. It undergoes dramatic changes along the stellar life, as shown by observational studies and also predicted by evolution models of angular momentum (Kawaler \cite{Kawaler88}; MacGregor \& Brenner \cite{MacGregor91}; Krishnamurthi et al. \cite{Krishnamurthi97}; Bouvier et al. \cite{Bouvier97}; Sills et al. \cite{Sills00}; Ivanova \& Taam \cite{Ivanova03}; Holzwarth \& Jardine \cite{Holzwarth07}). 
On one hand the properties of magnetic fields depend on rotation and mass, on the other hand, they play a fundamental role in altering the 
rotational properties of late-type stars. For example, they are responsible for the angular momentum loss and for the coupling mechanisms between  
the radiative core and the external convection zone (e.g., Barnes \cite{Barnes03}). Such an interplay between rotation and magnetic fields provides
us with a powerful tool to probe the stellar internal structure. \\
\indent 
A number of valuable ongoing projects (MONITOR, Hodgkin et al. \cite{Hodgkin06}; EXPLORE/OC, Extrasolar Planet Occultation Research, von Braun et al. \cite{von05}; RCT, Robotically Controlled Telescope project, Guinan et al. \cite{Guinan03}) are rapidly increasing our knowledge of the rotational
properties of late-type members of open clusters. However, the sequence of ages at which the angular momentum evolution has been studied still has significant  gaps, and the sample of periodic cluster members  is not as complete as necessary to fully constrain the various models proposed to describe the driving mechanisms of the angular
 momentum evolution. \\
   \indent
   We have selected  open clusters (Messina \cite{Messina07})  and young associations (Messina el al.  \cite{Messina09}) with an age in the range from 1 to 500 Myr  and, generally,  with no earlier rotation and magnetic activity studies. Top priority 
is given to the open clusters that  fill the gaps in the empirical description of the rotation/activity/age relationships. We have selected also a few extensively studied clusters such as the \object{Pleiades} and the Orion Nebula Cluster (Parihar et al. \cite{Parihar09}). The motivation is to make repeated observations  over several years to further enrich the sample of periodic variables and to explore the long-term magnetic activity, e.g., to search for activity cycles and surface differential rotation (SDR).
 Our sample also includes  open clusters  that were previously monitored  with different scientific motivations. The re-analysis of these archived  data time series 
can provide valuable results in the context of the RACE-OC project. That was the case of M37 (Messina et al. \cite{Messina08a}, hereinafter Paper I). Although it was initially observed to search for early-type pulsating variables (Kang et al. \cite{Kang07}), its data time series allowed us to determine, for the first time, the rotation period distribution of G stars at an age of about 550 Myr. Similarly, the M11 photometric data time series which were collected with the same goal (Koo et al. \cite{Koo07}), have allowed us to determine, again for the first time,
 the rotation period distribution of G stars at an age of about 230 Myr.
\begin{figure}
\begin{minipage}{10cm}
\centering
\includegraphics[scale = 0.5]{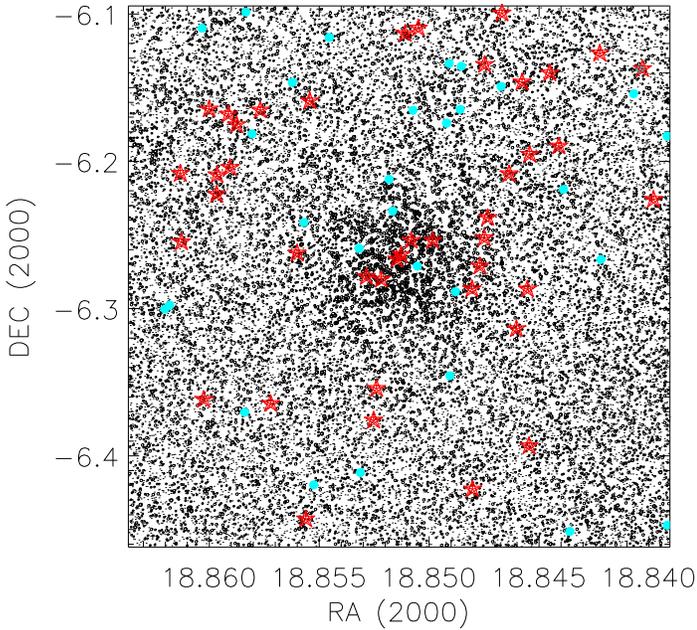}
\end{minipage}
\caption{\label{M11} \object{M\,11} field of view  (22.2$^{\prime}$$\times$ 22.2$^{\prime}$).  The newly discovered
and previously known periodic variables are plotted as  large red stars and light-blue bullets, respectively} 
\end{figure}

\object{M\,11} (NGC\,6705; RA$_{J2000.0}$ = 18:51:04, DEC$_{J2000.0}$ = $-$06:16:30) is a $\sim$230 Myr intermediate-age  open cluster at a distance of nearly $d$=2.0 kpc, (m$-$M)=12.69$\pm$0.1. The cluster is subjected to substantial reddening E(B$-$V)=0.428 mag  because of low galactic latitude ($b$=$-$2.8$^{\circ}$). It contains  thousands of members within an estimated radius of 16 arcmin. Gonzalez \& Wallerstein (\cite{Gonzalez00}) found a small metal excess, +0.10 dex, in agreement with the general trend of increasing metallicity with decreasing distance from the Galactic center.  In the following analysis we adopt the cluster parameters derived by  Sung et al. (\cite{Sung99}). \\
\indent
The first comprehensive variability study of M11 was carried out in 2002-2003  by Hargis et al. (\cite{Hargis05}) in the R band and on a 13.7$^{\prime}$$\times$13.7$^{\prime}$ field of view.  They discovered 39 variables, and for 32 of these they could determine the periodicity. Among late-type stars (B$-$V $ >$ 0.55), Hargis et al. (\cite{Hargis05})
 could identify   only 15  periodic eclipsing binaries,  probably because of  the large amplitude of light variation, but no periodic single stars.
The  poor seeing  ($\sim$ 5  arc-second), the use  of  a small
 telescope and the  faintness of the cluster late-type members  at a distance of
 about 2kpc  did not  allow them to  acquire sufficiently precise data  required to
detect very low amplitude variation due to spots. A second comprehensive variability study of M11 was carried out in 2004 by Koo et al. (\cite{Koo07}), who detected all the variables previously  found by Hargis et al. (\cite{Hargis05}) and 43 new periodic variables. Among late-type stars, they discovered  12 W UMa and 2 detached eclipsing binaries. Again, they did not report the discovery of any late-type low-amplitude periodic single stars, their study being focused on early-type pulsating variables.\\
\indent
Our investigation, which is based on  the Koo et al. (\cite{Koo07}) database, aims at detecting the periodicity of the  low-amplitude late-type single members of M11.
In Sect.\,2 we give details on observations and data analysis. The rotation period
search is presented in Sect.\,3, the results in Sect.\,4. Discussion and conclusions are given in Sect.5 and 6.

\section{Observations and data analysis}
The present study is based on observations taken in June 2004 with the 1.0m telescope at the Mt.\,Lemmon Optical 
Astronomy Observatory (LOAO) in Arizona (USA), which feeds a 2K$\times$2K CCD camera. The observed field of view (FoV) is about 22.2$^{\prime}$  
$\times$ 22.2$^{\prime}$ (see Fig.\,\ref{M11}) at the f/7.5 Cassegrain focus. We collected a sequence of 406 long- (600 s) and 595 short-exposed (60 s)
images  in the V-band filter  over a total time interval of 18 days. Additional
 observations  in the B bandpass  filter were  made on  one night  of October 2004, 
in order to construct  a  V vs. B$-$V color-magnitude  diagram.
A detailed description of these observations and data reduction can be found in Koo et al. (\cite{Koo07}) and Kim et al. (\cite{Kim01}).

\begin{figure}
\begin{minipage}{10cm}
\centerline{\psfig{file=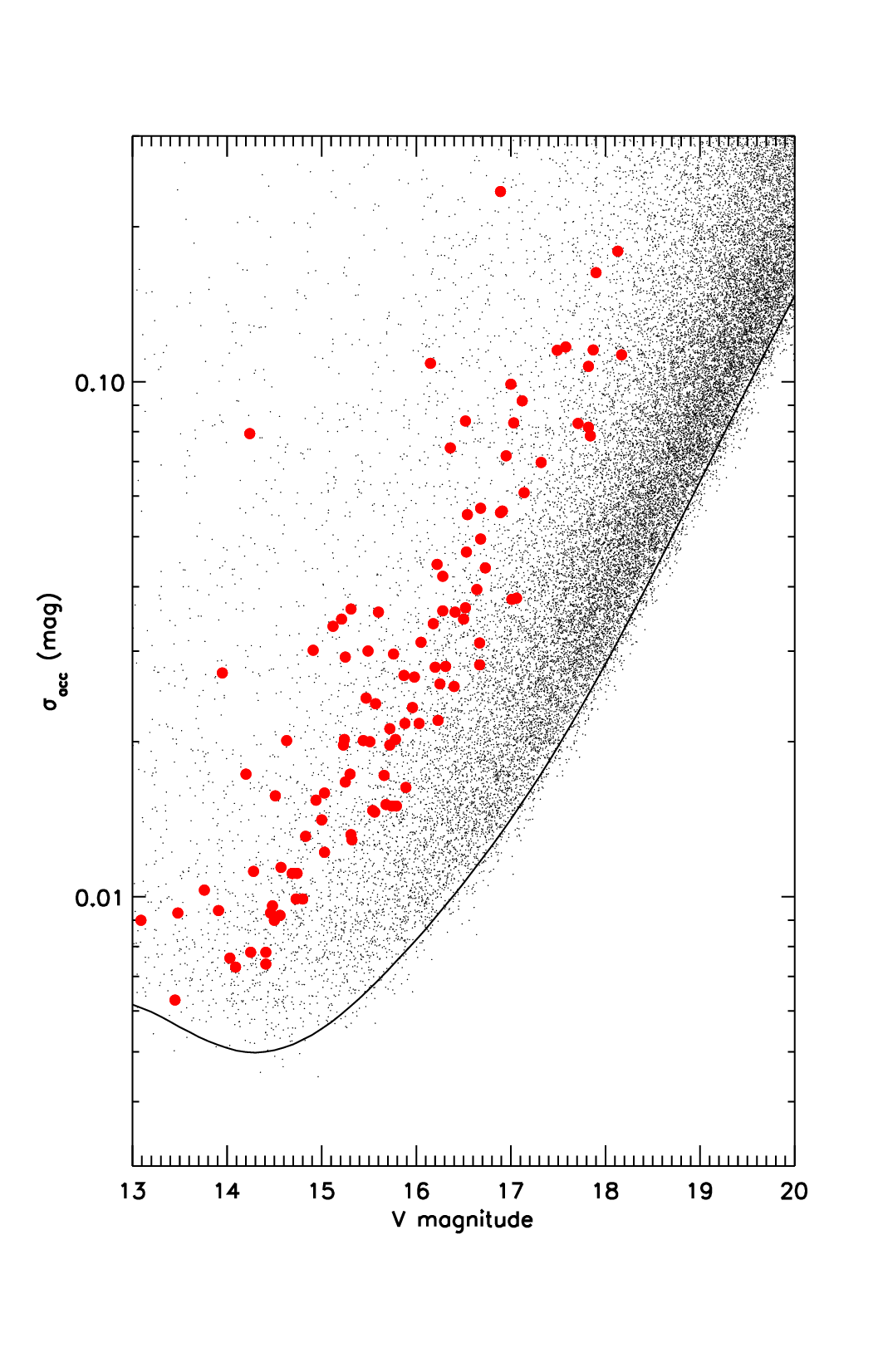,width=9cm,height=13cm,angle=0}
}
\end{minipage}
\caption{\label{accuracy} Photometric accuracy $\sigma_{\rm acc}$ vs. mean magnitude for the stars in the M11 FoV. 
The solid line represents a polynomial  fit to the distribution lower envelope. Small red bullets represent the periodic
 variable stars.}
\end{figure}
 We detected about 33400 stars in the 12 $<$ V $<$ 20 magnitude range in the  600-s long exposures.  Their coordinates, B and V magnitudes 
are available in the WEBDA open cluster database\footnote{http://www.univie.ac.at/webda/archive$_-$ngc.html}. These stars are represented in Fig.\,\ref{M11} with open circles. The symbol size is proportional to the star's brightness. The over-plotted large red stars and light-blue bullets represent the newly discovered  and already known periodic variables in the FoV under analysis.\\
\indent
Before analysing our magnitude time series for the rotation period search, we took care to eliminate possible outliers. First we disregarded all data points that deviated more than 4 standard deviations from the mean of the whole series data. Such a broad limit was adopted to include the light minima of  possible eclipsing binaries. In fact, we are aware from earlier studies of the presence of numerous eclipsing binaries in our FoV. Then, we computed a filtered version of the light curve by means of a sliding median boxcar filter with a boxcar extension equal to  1 hr.
This filtered light curve was subtracted from the original light curve and all the points deviating more than 3 standard deviations of the residuals were discarded. Finally, we computed normal points by binning the data on time intervals having the duration of about 1 hr, getting a light curve consisting of about 100 normal points on average. Each normal point is obtained by averaging about 4 consecutive frames collected within a time interval of 1 hr.
We adopted the average standard deviation of our normal points $\sigma_{\rm acc}$  as  photometric accuracy of our observations, rather
than the values computed by DAOPHOT while extracting the
PSF magnitudes. 
This standard deviation $\sigma_{\rm acc}$  is an empirical estimate of the effective precision of our photometry. 
It represents a conservative value because  the true observational  accuracy could be, in principle, even better for stars
showing substantial variability  within the  timescale  closer to our fixed binning time interval.
In Fig.\,\ref{accuracy} we plot $\sigma_{\rm acc}$ of the stars in the  $ 13.0 < $ V $ < 20.0$ magnitude 
range vs. their mean V magnitude. The lower envelope of the $\sigma_{\rm acc}$ distribution is populated by non variable 
and variable stars with least intrinsic variation. Such a  lower envelope gives a measure of the best photometric accuracy 
we achieved at different magnitudes.  In order to construct the relation between observational accuracy and magnitude,
we fitted a multi-order polynomial function to the lower $\sigma_{\rm acc}$  boundary. 
%, we followed the procedure given, e.g., by Roze \& Hintz 
%,(\cite{Roze07}), and fitted the lower envelope of  Fig.\,\ref{accuracy}  with a piecewise continuous function of the form:
%,\begin{tabular}{llr}
%,&\\
%,              & = $0.20 \times e^{-0.22V}$   &   12 $<$ V $\le 16$\\
%,$\sigma_{\rm acc}$ (mag) &\\
%,               & = $4.14 \times 10^{-8} e^{-0.735V}$ &    V $>  16$\\
%,
%,&\\
%\end{tabular}
The 600-s long exposures  have allowed us to achieve a photometric accuracy $\sigma_{\rm acc}$ in the V band of  about 0.006 mag  in the $ 13 < $ V $ < 16$ magnitude range, and better than 0.02 mag for all stars in the magnitude range  16.0 $<$ V $<$ 18.\\ 
\indent
We note that the photometric accuracy  is slightly poorer than the accuracy we
achieved for M37 (Paper I), although we have used the same telescope, instrumental setup, and reduction procedure.
 \object{M\,11} has a distance modulus about 1 magnitude larger than M37, is very rich, crowded and younger than M37. 
For example, our candidate periodic stars have on average 2 closeby stars within 5 arcsec and with a brightness difference down to 3 magnitudes.
However, we adopted  the PSF photometry which is the most accurate approach to determine the star's magnitude in crowded fields
and which should eliminate  any spurious flux contribution from closeby stars. Therefore, we 
suspect that the slightly poorer photometric accuracy of M11 with respect to M37 arises from its younger age and,
specifically, from the higher level of short-scale ($<$ 1hr) intrinsic variability.

\section{Rotation period search}

 We   have  used the
Scargle-Press method to search for significant periodicities related to the stellar rotation in
our data time series. In  the following  sub-sections we briefly  describe our
procedures to identify the periodic variables.
The period search was carried out on the 1-hr binned data time series, that is on
about 100 data points with respect to the original series of about 400 frames, and after discarding evident outliers, as discussed in the previous section.
\rm

\subsection{Scargle-Press periodogram}

The  Scargle technique  has  been  developed in  order  to search  for
significant  periodicities  in unevenly  sampled  data (Scargle  1982;
Horne \& Baliunas 1986). The algorithm calculates the normalized power
P$_N$($\omega$) for a given  angular frequency $\omega = 2\pi\nu$. The
highest   peaks  in  the   calculated  power   spectrum  (periodogram)
correspond to  the candidate periodicities  in the analyzed  data time series. 
 In order  to determine the significance level  of any candidate
periodic signal, the height of the corresponding power peak is related
with a false  alarm probability (FAP), that is  the probability that a
peak of given height is due to simply statistical variations, i.e.  to
Gaussian noise.  This method assumes  that each observed data point is
independent from the  others.  However, this is not  strictly true for
our data time series consisting of data consecutively collected within
the same  night and with  a time sampling  much shorter than  both the
periodic  or the  irregular  intrinsic variability  timescales we  are
looking  for (P$^d$=0.1-15).  The  impact of  this correlation  on the
period  determination  has  been   highlighted  by,  e.g.,  Herbst  \&
Wittenmyer (1996),  Stassun et  al. (1999), 
Rebull (2001),  Lamm  et al.   (2004). In
order to  overcome this  problem, we decided  to determine the  FAP in
different way than proposed by  Scargle (1982) and Horne  \& Baliunas
(1986),  the  latter being  only  based  on  the number  of  independent
frequencies.

\subsection{False alarm probability}
Following the approach outlined by Herbst et al. (2002),  randomized  time  
series  data  sets  were  created  by  randomly
scrambling the day number of  the Julian day while keeping photometric
magnitudes  and the  decimal part  of  the JD  unchanged. This  method
preserves  any correlation that exists in the  original  data set.  We
noticed  that Lamm et  al. (2004)  in order  to produce  the simulated
light  curves,  randomized the  observed  magnitudes,  instead of  the
epochs of  observation. Then, we  applied the periodogram  analysis to
about 10,000  "randomized'' data time series for each star.
We retained  the highest  power peak and  the corresponding  period of
each     computed    periodogram.   
The FAP related to a given power  P$_{ N}$ is taken as the fraction of
randomised  light curves that  have the  highest power  peak exceeding
P$_{ N}$ which, in turn, is the probability that a peak of this height
is simply  due to statistical  variations, i.e. white noise.   
The normalised power corresponding to
a FAP $=$ 0.01 was  found to be P$_{\rm N}$= 9.2. 

\begin{figure}
\begin{minipage}{10cm}
\includegraphics[scale = 0.4, trim = 0 0 0 0, clip, angle=90]{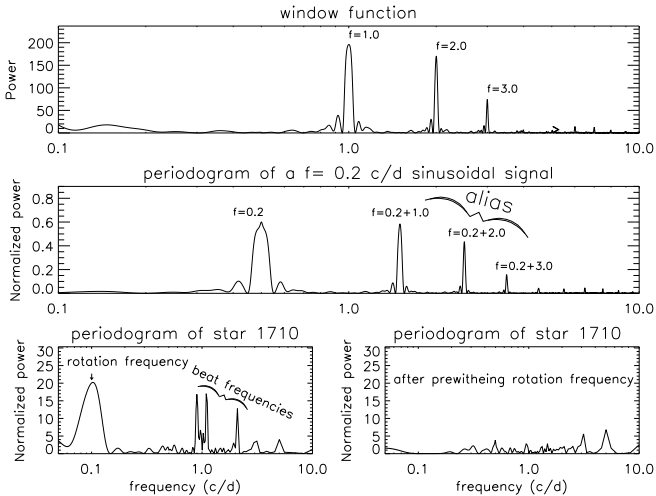}
\end{minipage}
\caption{\label{window} \it Top panel\rm: typical window function deriving from the data sampling and
total  time interval  of the  M11 observations.  \it Middle panel\rm: aliasing effect of the window function on an example periodic 
signal with frequency f=0.2 cd$^{-1}$.  \it Bottom left panel\rm: periodogram of star 1710 with rotation frequency together with its beat frequencies.
 \it Bottom right panel\rm: disappearing of beat frequencies after prewhitening the rotation frequency.}
\end{figure}

\begin{figure*}
\begin{minipage}{18cm}
\includegraphics[scale=0.9, trim = 0 500 0 0, clip, angle=0]{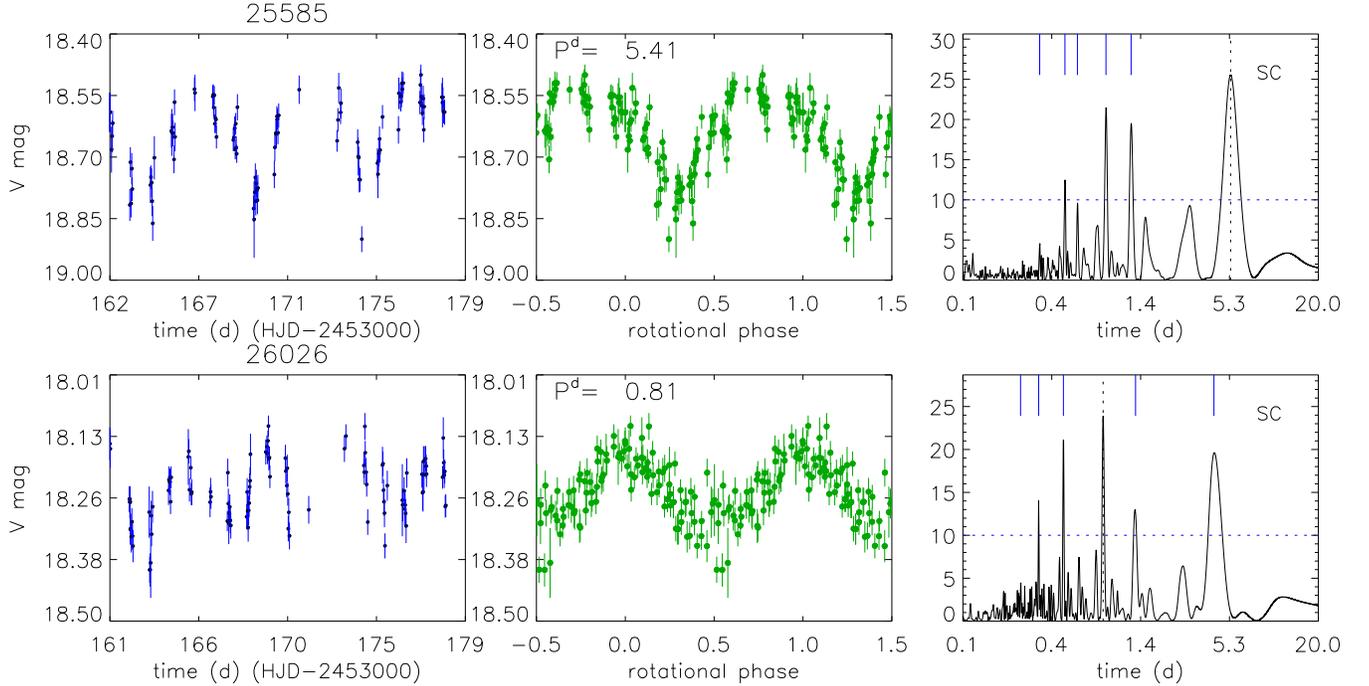}
\end{minipage}
\caption{\label{example} \it From left to right panel: \rm V-band data time series with uncertainties; folded light curves with phases computed by using the derived period; Scargle  normalized periodogram. The vertical dashed line indicates the major peak related to the rotation period, whereas the solid top bars indicates its beat periods.The horizontal dotted line indicates the 99\% confidence level.}
\end{figure*}

\subsection{Alias detection}

In order to identify the true periodicities in the periodogram, it is
crucial to take into account that a few peaks, even with large power, are aliases  
arising from both the data sampling and the total duration of the observation run.

An inspection of the spectral window function helps to identify which
peaks  in the periodogram may be alias.

In the top panel of Fig.\,\ref{window} we plot an example of the spectral window function
of our data. Due to the sampling interval of about 1 day imposed by the rotation of the Earth 
and the fixed longitude of the observation site, the window function has major peaks at  $f$ = $\pm$$n$ cd$^{-1}$ ($n$ integer), together with 
smaller sidelobes in the range $f$ = 1 $\pm$ 0.1 cd$^{-1}$ around the major peak. 

In the middle panel of Fig.\,\ref{window} we plot the effects of the spectral window function on an example  strictly periodic 
signal of $f$ = 0.2 cd$^{-1}$ with the same data sampling as the real data. Apart from the power peak corresponding
to the true periodic signal at f = 0.2 cd$^{-1}$,
a number of alias peaks appear as consequence of the convolution between the power spectrum
and the window function. All these alias periods are beat periods (B) between the star's rotation period (P) and the data sampling and they obey to the relation
\begin{equation}
\frac{1}{B} = \frac{1}{P} \pm n \,\,\,\,(n=1,2,3,...)
\end{equation}

A method  to check whether peaks at short periods are beat periods is to perform a prewhitening of the data time series
by fitting and removing a sinusoid with the star's rotation period from the data.

In the bottom left panel of Fig.\,\ref{window} we plot the case of  star 1710 whose periodogram shows
four peaks with  confidence level larger than 99\%. Assuming that the highest peak at about f=0.09 cd$^{-1}$
indicates the star's rotation frequency, we see on the bottom right panel of Fig.\,\ref{window} that, after removing the primary frequency from the data time series and recomputing the periodogram, actually all the other peaks disappear, confirming
they are actually beat frequencies.

\subsection{Uncertainty with the rotation periods}

In order to compute the error associated with the periods
we followed the  method used by Lamm et al.  (2004) where
the uncertainty  can be written as
\begin{equation}
\Delta P = \frac{\delta \nu P^2}{2}
\end{equation}
$\delta\nu$ is  the finite  frequency resolution  of  the power
spectrum and is equal  to the full width at half maximum of  the main peak of
the window  function w($\nu$).   If the time  sampling is not  too non-uniform,
 which is the case  related to our observations, then $\delta\nu
\simeq 1/T$,   where T is the total  time span of  the observations. From
Eq.\,(2) it  is clear that  the uncertainty  not
only depends  on the frequency resolution  (total time  span) but is also
proportional to  the square of the  period. We computed the error
on  the period  following also the prescription suggested by  the Horne \&
Baliunas  (1986),  which  is  based on the formulation  given  by  Kovacs
(1981). The uncertainty computed according
to  Eq.\,(2) was found  to be  factor 5-10  larger than  by the  technique of Horne \& Baliunas  (1986). In this paper
we report the error computed with  Eq.\,(2).
Hence, it can be considered as  an upper limit, and the precision in the
period could be better than that we quote in this paper.

\section{Results}
Our period search allowed us to  detect 75 new \rm  periodic variables in the 22.2$^{\prime}$$\times$22.2$^{\prime}$ FoV centered in M11. The results are summarized in 
Table\,\ref{tab_period} where we list the following information: internal Identification Number (ID);
Webda identification number; coordinates (J2000);  periodicity (P) and its uncertainty ($\Delta$P); normalised peak power (P$_{\rm N}$);  average V magnitude ($<$V$>$); dereddened (B$-$V)$_0$ color;  photometric accuracy ($\sigma_{\rm acc}$); light curve amplitude ($\Delta$V). The latter is computed by making the difference between the median values of  the upper and lower 15\%   light curve normal points (see, e.g., Herbst et al. \cite{Herbst02}). That prevents  overestimation of  the amplitude due to possible residual outliers. We list a note about the membership probability to the cluster, as will be discussed in the next section: 'a' is for high-probability members; 'b' for lower-probability members; 'n' for non-members. Finally, we indicate with a flag 'y' the eclipsing binary systems as classified on the basis of the shape of their light curve.
 In Fig.\,\ref{example} we plot, as an example, the results of period search for the stars 25585 and 26026. From left to right panels we plot the V-band magnitude vs. time with overplotted the uncertainties; the phased light curve, which is folded by using the derived period; the Scargle periodogram, where the major power peak related to the observed periodicity is marked by a vertical dotted line, beat periods by solid lines, and the 99\% confidence level by dotted horizontal line.\\ 
\indent
The complete set of 75 light curves, together with Scargle periodograms, are plotted in the online Figs.\,\ref{a1}-\ref{n8}. \\
%\bf We note that the light curve amplitude of the proposed periodic variables is at least a factor 2 larger than the photometric accuracy
%and at least a factor 3 larger than the average dispersion of normal points with respect to the sinusoidal fit. Both circumstances
%give support to the correctness of rotation period determination. \rm

\indent

\section{Discussion}
Our major aim is to determine the rotation period distribution and the activity level of the late-type
members of M11. Therefore, we have first to identify the cluster members among the 75 newly discovered  
periodic variables, and subsequently to select only  the late-type members whose discovered periodicity is
likely related to stellar rotation. We will exclude
evident eclipsing binaries from the final sample of periodic late-type members. 
In fact, due to tidal synchronization, the rotational evolution of the components 
of close binary systems significantly differs from the evolution history of single stars, 
on which we are focussed.

\subsection{Candidate member selection}

\begin{figure*}
\begin{minipage}{18cm}
\centerline{
\psfig{file=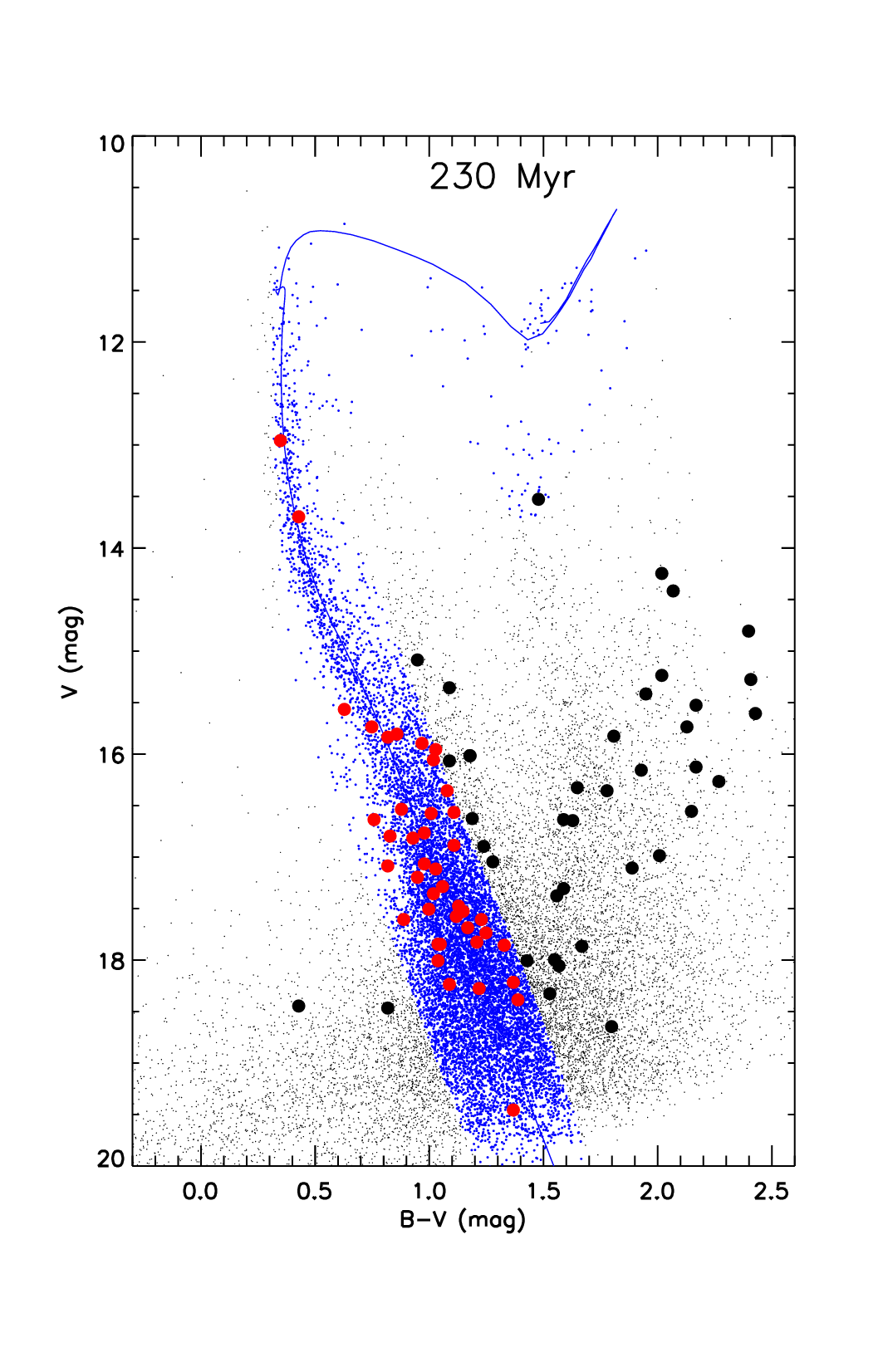,width=3.5cm,height=13cm,width=9cm,angle=0}
\psfig{file=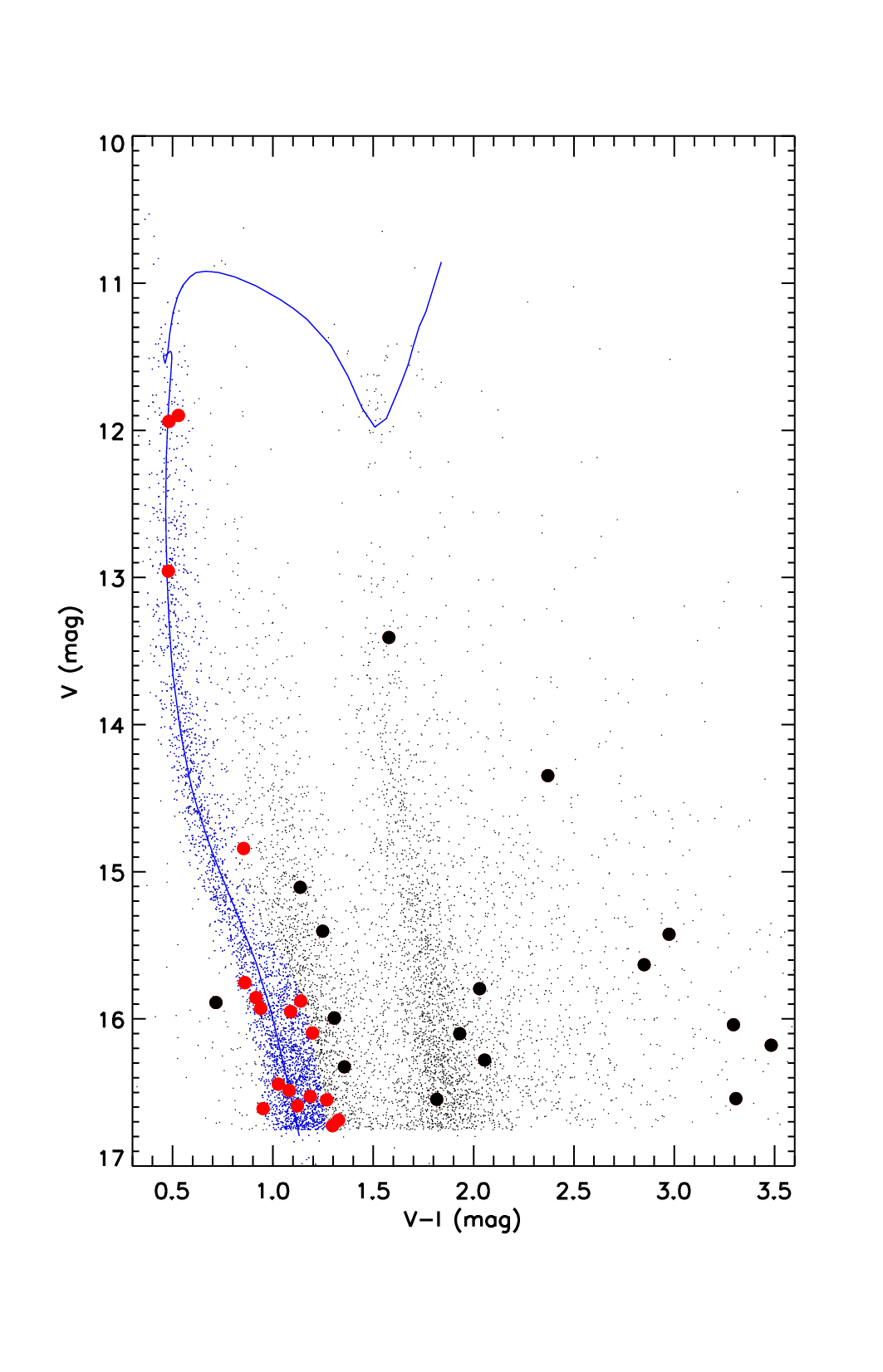,height=13cm,width=9cm,angle=0}}
\end{minipage}
\caption{\label{HR} \it Left panel: \rm V vs . B$-$V diagram of the stars (dots) detected in the 22.2$^{\prime}$$\times$22.2$^{\prime}$ field of \object{M\,11}.  Photometric candidate members are plotted with blue small bullets, newly-discovered periodic candidate members are plotted with red large and black bullets. The solid line represents the isochrone corresponding to an age of  log t=8.35, E(B$-$V) = 0.428 mag, (m$-$M)=12.69 mag (Sung et al. \cite{Sung99}).  \it Right panel:\rm  same as left panel but with V$-$I colors from Sung et al. (\cite{Sung99}). Note that neither V nor V$-$I values are available for stars fainter than $\sim$ 16.7.}
\end{figure*}

M11 has  a very rich stellar background and  it is a very difficult task to identify the cluster members.
The proper motion  studies carried out  by McNamara et
al. (1977), Su  et al. (1998), and Dias et al. (2006)  provide information  only on  the
bright  (V$<$  16 mag) members. The radial velocity  survey carried
out by Mathieu et al. (1986) is also limited to a handful of stars.
 Our proposed periodic variables have information neither on proper motion nor on radial velocity. 
So we have left no option other than using the following two criteria to identify cluster members:
\it i) \rm the photometric membership, that is the distance of any star from the theoretical isochrone
in the color-magnitude diagram (see, e.g., Irwin 2006, 2008); \it ii) \rm the spatial distance from the cluster center.\\
\indent
In Fig.\,\ref{HR} we plot the color-magnitude V vs. B$-$V  (left panel) and V vs. V$-$I (right panel) diagrams of all stars (small dots) detected in our FoV. The B$-$V colors are ours, whereas the V$-$I colors are taken from Sung et al. (1999) and available only for stars brighter than V$\simeq$ 16.7.
We overplot the color-reddened 230-Myr isochrone derived from Girardi et al. (2000), and corresponding to the cluster parameters taken from  Sung et al.  (1999).\\
\indent
In the first criterion, we assign two probability levels on the basis of the distance from the isochrone: 
'1a' to stars with distance smaller than $\pm$(1.4+$\sigma$) mag (blue small dots in Fig.\,\ref{HR}), '1b' to stars with distance larger than $\pm$(1.4+$\sigma$) mag (black small dots) in the case of V vs. B$-$V diagram. Similarly, we assign  '2a' and '2b' in the case of the  V vs. V$-$I diagram.
Red bullets represent the proposed periodic variables with probability membership '1a' (left) and '2a' (right), whereas
black bullets represent the proposed periodic variables with probability membership '1b' and '2b', respectively.
The quantity $\sigma$ takes into account the increasing photometric uncertainty towards fainter stars.\\
\begin{figure}
\begin{minipage}{10cm}
\psfig{file=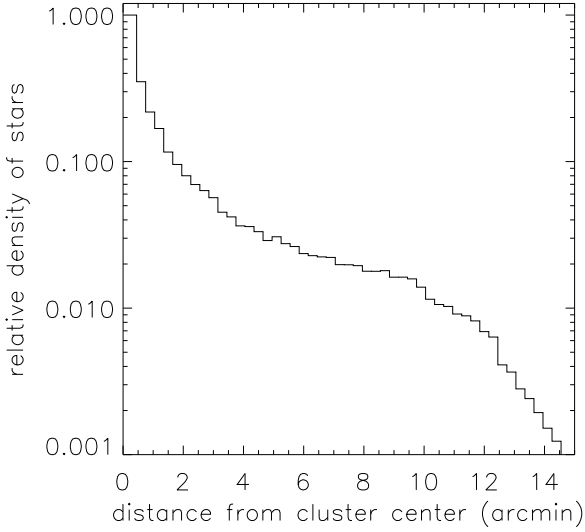,height=8cm,width=8cm,angle=0}
\end{minipage}
\caption{\label{spatial} Relative density of stars in the M11 FoV with respect to the cluster center.\rm}
\end{figure}
\begin{figure}
\begin{minipage}{10cm}
\psfig{file=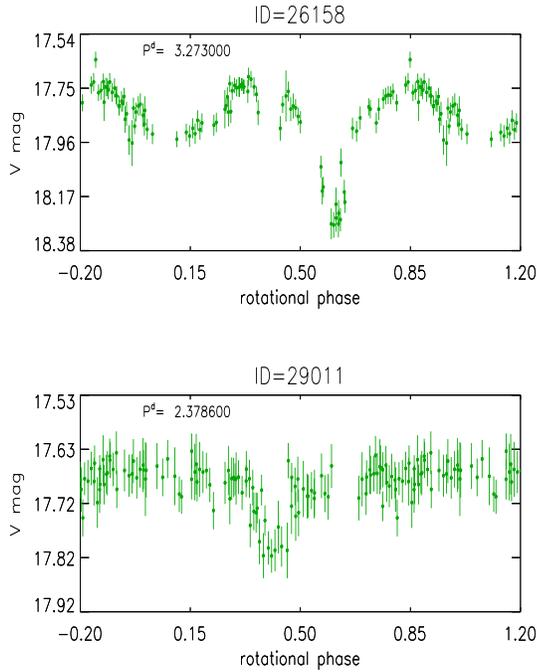,height=16cm,width=9cm,angle=0}
\end{minipage}
\vspace{-5cm}
\caption{\label{eclipsing} Eclipsing binaries newly discovered among periodic members of M11 .\rm}
\end{figure}
\indent
In the second criterion, we assign two probability levels on the basis of the projected angular distance from the cluster center and the relative density 
distribution shown in Fig.\,\ref{spatial}: 
'3a' to stars with distance smaller than 10$^{\prime}$, '3b' to stars with distance larger than  10$^{\prime}$.\\
\indent
As a final result,  we assign a high-probability flag 'a' to stars having contemporarily  '1a', '2a', and '3a'; a lower-probability flag 'b' to stars having '1a', '3a' and no measured V$-$I color. All the remaining stars are classified to be non members 'n' and are excluded from the analysis of  the cluster rotation period distribution. The probability flags are listed in Table\,\ref{tab_period}.\\
\indent
Our final sample of newly discovered variables consists of: 9 high-probability members (4 early-type stars with (B$-$V)$_0$$<0.5$, 5 late-type stars with (B$-$V)$_0$$\ge 0.5$); 29 lower-probability members (2 early-type, 25 late-type, 2 eclipsing binaries); 37 non members which will not considered in the following analysis. In Fig.\,\ref{eclipsing} we plot the 2 newly discovered  eclipsing binaries cluster members.
\rm
\subsection{Rotation period distribution}

\begin{figure*}
\begin{minipage}{18cm}
\centerline{
\psfig{file=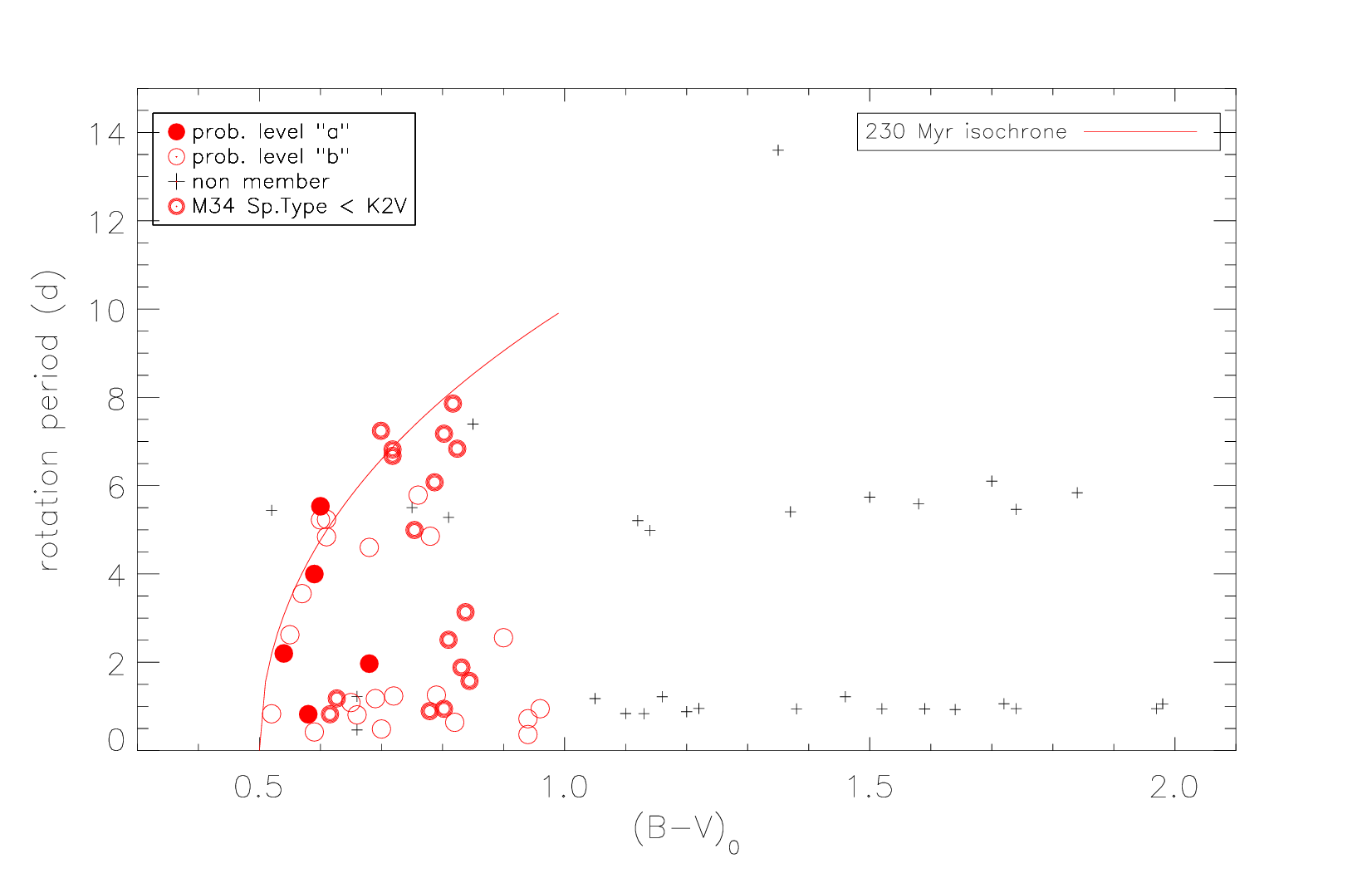,width=9cm,height=6.0cm,angle=0}
\psfig{file=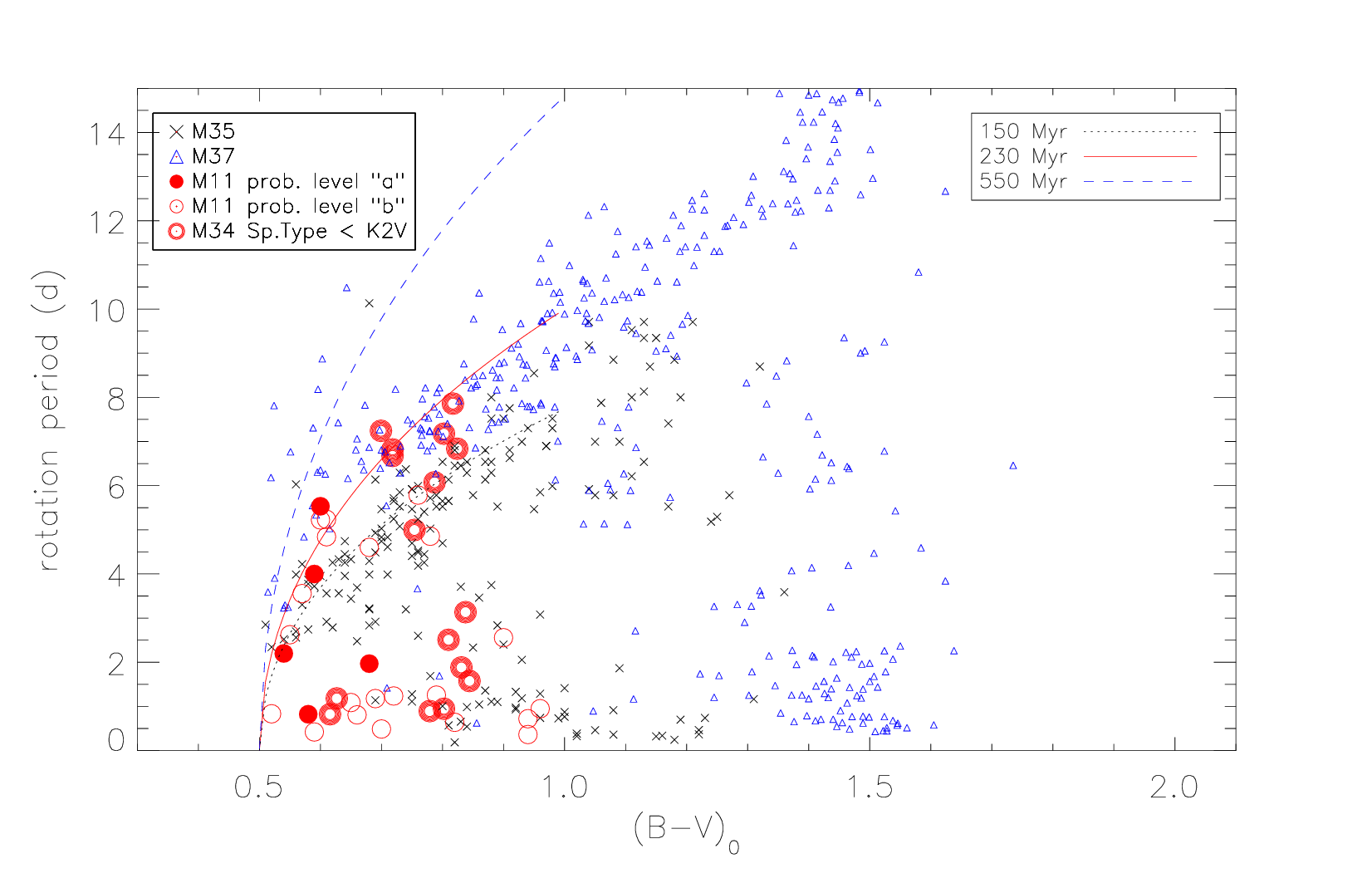,width=9cm,height=6.0cm,angle=0}
}
\end{minipage}
\caption{\label{gyro} \it Left panel\rm: Rotation period vs. (B$-$V)$_0$ colour for the members of  \object{M11}. Different symbols are used to represents high- and low membership probability. The periods of the M34 members  with spectral type earlier than K2/3 are also plotted. The solid curve represents the gyro-isochrone at a nominal age of 230 Myr.  \it Right panel\rm: same as left panel, but with overplotted the rotation period distribution of \object{M37} (triangles) and \object{M35} (asterisks). The family of age-parameterised curves from gyrochronology corresponding to ages of 150, 230 and 550 Myr is overplotted.}
\end{figure*}

As anticipated, our attention is focussed on the late-type members of \object{M\,11}. In fact,  the detected periodicity of these late-type stars most likely represents the stellar rotation period. The  variability observed  over time scales from several hours to days arises from non-uniformly distributed, cool-spotted regions on the stellar photosphere, which are carried in and out of view by the star's rotation. 
We excluded from our analysis all periodic variables bluer than (B$-$V)$_0$=0.5, whose periodic variability is likely related to pulsations. We excluded also the already known (see Koo et al. \cite{Koo07}) as well as the newly discovered W UMa-type and detached eclipsing binaries, their angular momentum evolution being altered by tidal synchronization between stellar components.\\
\indent
All periodic members discovered by us have (B$-$V)$_0$$<$0.9, which roughly corresponds to main sequence stars of spectral type earlier than K2/3 and mass  M/M$_{\odot} \ga 0.8 $. Therefore, our analysis has provided for the first time the rotation period distribution of the G-type and early-K members of M11. The absence of periodic middle-K or later-type stars must be ascribed to the lower photometric accuracy, which becomes comparable to or larger than the periodic variability amplitude due to starspots for the fainter stars.\\
\indent
In Fig.\,\ref{gyro} we plot the rotation period distribution of late-type (bona-fide) single stars  vs. the (B$-$V)$_0$ color. Different symbols are used to plot stars with different levels of membership probability to the cluster, according to the criteria outlined in the previous Section. Specifically, 5 stars are found to be high-probability members ('a') and plotted with filled bullets; 16 stars have a lower membership probability ('b') and are plotted with open circles. Here we like to remind that such lower probability level of membership arises from 
unavailability of V$-$I measurements (see right panel of Fig.\,\ref{HR}). In the following analysis we consider all together, although plotted with different symbols, both stars with  'a' and 'b' membership probability. \\
\indent
A cluster of  slightly younger age (about 200 Myr) and with known rotational properties is M34 (Irwin et al. \cite{Irwin06}). The survey of Irwin et al. (\cite{Irwin06}) allowed to
determine the rotation period distribution of the lower-mass K-M stars. In  Fig.\,\ref{gyro} we plot with partially filled-in bullets
the 16 periodic variables of M34 with spectral type earlier than K2. The original (V$-$I)$_0$ colors in Irwin et al. have been transformed into (B$-$V)$_0$ colors by using standard stars color-color relations (Cox 2000). These additional data  allow us to get a more numerous sample of periodic stars and, consequently, to derive a reliable rotation period distribution  of G0-K2 stars at an age of about 200-230 Myr.\\
\indent
As observed in other young open clusters, also the M11 members are found to distribute between two different rotation regimes. There is a fraction of stars whose
rotation period is larger than about 1-2 days. Their distribution displays an upper envelope which increases with increasing B$-$V color. These stars belong to the so called 'interface sequence', according to the classification scheme of Barnes (Barnes \cite{Barnes03}). These stars have experienced the effects of rotation braking by stellar magnetized winds, and are expected to continue slowing down to reach, by an age of about 500-600 Myr, a one-to one dependence between rotation period and color (see, e.g., Collier Cameron et al. 2009).
A second fraction of stars with period smaller than  1-2 days belongs to the so called 'convective sequence'. They have not experienced yet significant spin down and are progressively moving towards the interface sequence of slow rotators. Bluer stars in this sequence  leave first and the B$-$V color at which the sequence begins varies with the cluster age. The solid line is the theoretical curve, computed by using Eq.\, (1) of Barnes (2003) - P=$\sqrt(t)$ f(B$-$V) where f(B$-$V) represents the color dependence of the data -. It represents the expected upper envelope of the  rotation period distribution at a nominal age of  $\sim$ 230 Myr, assuming  a rotation rate decay according to the Skumanich law (Skumanich \cite{Skumanich72}).
We refer the reader to Barnes (\cite{Barnes03}) for a detailed description of both sequences. The agreement between the observed interface sequence and its expected distribution is quite good and gives support to the estimated cluster age.\\
\indent
From our analysis we find that the G-type members of M11 have a median rotation period of P=4.8 days. This is the first 
determination of median rotation period of G-type stars at an age of 230 Myr, which is the only available value in the age range 
from 150 Myr (M35) to 550 Myr (M37).\\

\subsection{Angular momentum evolution}
The variation vs. time of the median rotation period of cluster members is the quantity generally used to detect and investigate 
the rotational evolution of low-mass stars. \\
\indent
In the right panel of Fig.\,\ref{gyro}, we compare the rotation period distribution of M11 with the distributions of the 
younger open cluster M35 (150 Myr; Meibom et al. \cite{Meibom09})  and of the older open cluster M37 (550 Myr; Hartman et al. \cite{Hartman09}; Messina et al.\,\cite{Messina08a}). 
Focussing our attention on the G-type slow rotating stars of these clusters, that is on stars in the interface sequence and in the 0.55 $<$ (B$-$V)$_0$ $<$ 0.9 colour range,
we find that  the median rotation periods are P$_{\rm M35}$=4.4d, P$_{\rm M11}$=4.8d, P$_{\rm M37}$=6.8d. This is the major result of our analysis: we have determined the median rotation period of G stars at an age of 230 Myr, at which to date no information on rotational properties was available. Moreover, the comparison with the median period of two other clusters shows that the value we found is consistent with the expected rotation slow down with age.\\
We  plot the age-parameterized family of  theoretical curves corresponding to nominal ages of  $\sim$ 150 Myr (black dotted line)  $\sim$ 230 Myr 
(red solid line), and $\sim$ 550 (blue dashed line). The younger gyro-isochrones (150 and 230 Myr) appear to fit the observed distribution upper envelopes
 better than the older gyro-isochrone at 550 Myr. However, this discrepancy depends on the parametric curve rather than on the age of M37, which is quite well established (see Hartman et al. \cite{Hartman09}).
Two important results arise from comparing data with isochrones. First, the agreements shows the validity of the Skumamnich rotation rate decay, that is a dependence of rotation on square root of time. Second, the Barnes parameterization very well reproduces the observed rotation period distribution 
at an age of about 230 Myr.\\
\indent
Now we turn our analysis to the fast rotators. In this case we note some sort of discrepancy with respect to the younger cluster M35.
In fact, in M35 the convective sequence begins approximately at (B$-$V)$_0$=0.8. We expect in older clusters such sequence to begin
at redder colors. On the contrary, we note in both M34 and M11 a number of  fast rotating stars bluer than(B$-$V)$_0$$<0.8$.
Actually, this discrepancy may be explained without need of contradicting the rotation evolutionary sequence described above.
Firstly, we see that these unexpected blue fast rotators have lower membership probability.
 As discussed by Sung et al. (\cite{Sung99})  the cluster is  at low galactic latitude ($\sim-3^{\circ}$), 
near the Scutum star cloud and  the Sagittarius-Carina arm, resulting in very large contamination of the cluster color-magnitude diagram 
 from the relatively young  (and, therefore, fast rotating) field population. Therefore, these blue fast rotators  may just  be interlopers.
 Alternatively, some of these fast rotators may be cluster members,  but they may belong to close binaries of BY Dra, RS CVn, W UMa or FK Com types,
which are not identified as such variables, yet, and are rapidly rotating due to tidal interaction and synchronisation.
Indeed, all the already known and newly discovered binary stars with (B$-$V)$_0$ $>$ 0.5 and with high-probability membership,   if plotted in Fig.\,\ref{gyro}, would all populate
the convective sequence.
A third possibility may be the  false  detections of the periods. For example, the presence of two activity centers in the stellar photosphere at the epochs of observations, about 180$^{\circ}$ away in longitude from each other, can produce a light modulation with half of the true rotation period. In these cases, consecutive observation seasons are needed to detect the true rotation period (Parihar et al. \cite{Parihar09}). \\
\indent
%This is the second paper of the RACE-OC series. In the first we presented for the first time the rotation period distribution of G stars at an age of about 550 Myr (M37).
%A similar rotation period distribution in M37 (but extended to lower mass stars) was determined independently by Hartman et al. (\cite{Hartman09}). With the present result on G stars at an age of about 230 Myr we have now the possibility to better describe the rotational history of G stars in the age interval between Pleiades and Hyades stars.

\subsection{Photospheric activity}
\begin{figure}
\begin{minipage}{10cm}
\centerline{
\psfig{file=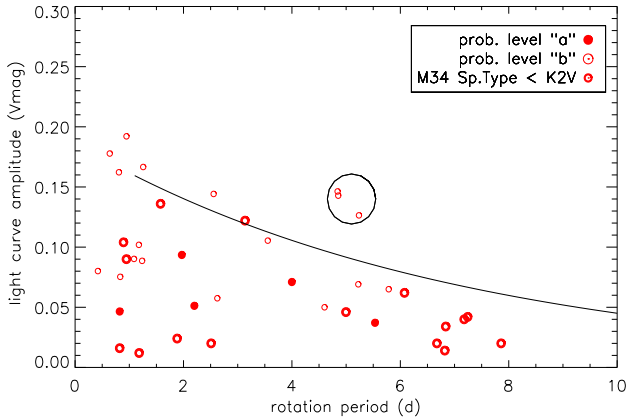,width=9cm,height=6.0cm,angle=0}
}
\end{minipage} 
\caption{V-band light curve amplitude ($\Delta$V) vs. rotation period for the same stars plotted in the left panel of Fig.\,\ref{gyro}. The solid line represents the upper envelope
of the $\Delta$V vs. rotation period  distribution of dwarf  G  stars  from Messina et al. (\cite{Messina03}), whereas the circled three stars above such boundary are suspected
binary stars.\label{amplitude}}
\end{figure}

The photometric variability we observed in the proposed late-type periodic members of M11 arises from the presence of cool photospheric spots whose visibility is modulated by the stellar rotation. The total amount of cool spots is related to surface magnetic fields, whose filling factor depends on the rotation rate and on the depth of the convection zone (or mass). The amplitude of the observed variability, specifically the peak-to-peak light curve amplitude, is suitable to trace the
 dependence of the magnetic field filling factor on rotation and mass. In Fig.\,\ref{amplitude} we plot the amplitude of the V-band light curve of the M11 late-type periodic members together with the  M34 members. The symbols have same meaning than in Fig.\,\ref{gyro}. We see that the upper envelope of the light curve amplitude decreases with increasing rotation period. This is consistent with the expected dependence on rotation rate of the efficiency of magnetic filed generation and intensification by an $\alpha\Omega$ dynamo. 
The solid line represents the fit to the upper envelope of the light curve amplitude distribution of slow rotating Pleiades G stars (110 Myr) taken from  Messina et al. (\cite{Messina01}, \cite{Messina03}). With the data at our disposal, we do not see any significant difference between the 110-Myr and the 230-Myr
distribution upper envelopes. Our sample of only G stars does not allow us to say anything about the mass dependence of the amplitude-rotation relation.\\
As discussed in Paper I, there is some marginal (yet) evidence of the existence of some other age-dependent quantity, in addition to mass and rotation, 
that controls the level of (photospheric at least) activity,  and makes older stars less active than younger stars (see also Messina et al. 2009). A similar suspect was already raised by Messina et al. (2001), who found evidence that, for a fixed mass and rotation period, the level of starspot activity increases (or alternatively the gross surface distribution of spots changes)  from the zero-age main sequence up to the \object{Pleiades} age ($\sim$120 Myr) and then  it decreases  with age. With the data of M11 G stars in our hand we can state that up the the M11 age  the level of activity remains at highest levels. On the contrary, the K-type stars at an age of  200 Myr already show a significantly decreased activity with respect to the younger K-type \object{Pleiades} stars, as shown in Fig.\,16 of the paper by Irwin et al. (\cite{Irwin06}). \\
\indent
It is interesting to note that, when we select the final sample of periodic variables to determine the rotation period distribution, the existence of a rotation-activity relation provides an independent  tool to identify either possible binary systems or erroneous period determination. In fact, due to eclipse minima and enhanced rotation with respect to single stars, close binaries show light curve amplitudes much larger than observed in single stars. Similarly, an incorrect period determination,  specifically an incorrect  value larger than the true rotation period (e.g. its beat period), produces an outlier in the rotation-activity distribution. Such outliers allow us
to identify, or at least to defer  for additional study, stars which may contaminate the final sample. In our case, the stars 10861, 18776 and 22541 (which are circled  in Fig.\,\ref{amplitude}) show a light curve amplitude larger than ever observed in stars of similar mass and rotation period. Follow-up observations may reveal that the adopted period is actually the (long) beat period. Similarly, this method can allowus to identify, and consequently exclude from the final sample,   W-UMa close binaries whose light curve amplitudes  clearly deviate from the distribution upper envelope and whose light curve shape are much scattered to identify its W UMa nature.

\section{Conclusions}
We have analysed V-band magnitude time series of the young 230-Myr open cluster M11 collected in 2004 at the LOAO observatory. 
Our analysis has allowed us to derive the following results:

\begin{itemize}
\item We have discovered 75 new periodic variables in a 22.2$^{\prime}$x 22.2$^{\prime}$ FoV centered in M11. Considering that 64 periodic variables were known from earlier studies, the total number of known periodic variable in this FoV is 139.

\item 2 newly discovered periodic variable members are  likely eclipsing binary systems. The discovery of such systems,  whose absolute physical parameters can be determined with  spectroscopic and multiband photometric follow-up studies, is relevant to better determine the distance to the cluster as well as to the understanding of the dynamical evolution of binary systems at intermediate age.

\item Out of 75 new periodic variables, we have found that 30 stars are likely single late-type periodic cluster members. 
Such stars have colors in the range 0.5$<$(B$-$V)$_0$$<$0.9, which corresponds to main sequence spectral type earlier than K2 and
stellar  mass larger than 0.8 M$_{\odot}$.

\item By adding to our data 16 G-type periodic variables 
belonging to the almost coeval cluster M34 and taken from the literature, we have determined from a sample of 46 stars the median rotation period P=4.8d of the G stars at an age of about 200-230 Myr for the first time.  In fact, the rotational properties of G stars in the age range 150-550 Myr were unknown before the present study.

\item
The median period we determined is longer than the median period P=4.4d of G stars in the younger cluster M35+M34 and shorter than the median period P=6.8d of G stars in the older cluster M37. This result is in good agreement with the expected scenario of rotation spin down with age according to the 
Skumanich rotation decay law.

\item The distribution of the light curve amplitude of G-type members of M11 is found to decrease with increasing rotation period. This activity-rotation relation is found either in other open clusters or in field stars and supports the expected operation of an $\alpha$$\Omega$ dynamo whose efficiency is related to both rotation and convection zone depth. The average level of acitvity of these 230Myr old stars is found to be comparable to the activity level of the younger Pleiades stars.
\end{itemize}

\begin{acknowledgements}
This work was supported  by the  Italian Ministero dell'Universit\`a, Istruzione e Ricerca (MIUR) and the National Institute for Astrophysics (INAF).
The extensive  use of  the SIMBAD  and ADS databases  operated   by  the  CDS  center,   
Strasbourg,  France,  is gratefully acknowledged. 
This research has made use of the WEBDA database, operated at the Institute for Astronomy of the University of Vienna. We thank 
the Referee for very useful comments and suggestions.
\end{acknowledgements}

\clearpage

\begin{deluxetable}{rrcccc@{\hspace{.1cm}}c@{\hspace{.1cm}}c@{\hspace{.1cm}}ccccc@{\hspace{.1cm}}c@{\hspace{.1cm}}}
\tablecolumns{12}
\tablecaption{\label{tab_period} Summary of period search: Star's internal identification number (ID); Webda number; Right Ascension and Declination; Period  (P) and and its uncertainty ($\Delta$P); Normalised peak power (P$_{\rm N}$); Average V magnitude ($<$V$>$); Dereddened B$-$V color; Photometric accuracy ($\sigma_{\rm acc}$), Light curve amplitude ($\Delta$V);  Membership probability; note on binarity based on light curve shape. }
\tablewidth{0pt}
 \tablehead{
\colhead {ID}   & \colhead {ID} & \colhead {RA} & \colhead {DEC} & \colhead {P$\pm\Delta$P} & \colhead {P$_{\rm N}$}  & \colhead {$<$V$>$} & \colhead {(B$-$V)$_0$} & \colhead {$\sigma_{\rm acc}$ } & \colhead {$\Delta$V } & \colhead {Mem}   & \colhead {Binary}\\   
\colhead {}   & \colhead {Webda} & \colhead {(hh:mm:ss)} & \colhead {(dd:mm:ss)} & \colhead {(d)} & \colhead {}  & \colhead {(mag)
} & \colhead {(mag)} & \colhead {(mag)} &  \colhead  {(mag)} & \colhead {prob.} & \\   
}
\startdata
\hline
185   &  241  &  18:51:36.204  &  -06:16:40.24  &  0.94  $\pm$  0.01  &  15.98  &  13.91  &  1.59  &  0.0094  &  0.0602  &  n     &  \\
195   &  253  &  18:51:35.597  &  -06:09:11.90  &  0.466  $\pm$  0.003  &  16.73  &  14.74  &  0.66  &  0.0111  &  0.0483  &  n       &  \\
227   &  290  &  18:51:33.843  &  -06:20:27.27  &  0.94  $\pm$  0.01  &  21.06  &  14.09  &  1.52  &  0.0073  &  0.1116  &  n      &  \\
414   &  493  &  18:51:22.576  &  -06:21:51.37  &  0.249  $\pm$  0.001  &  15.08  &  14.24  &  0.20  &  0.0793  &  0.1723  &  b      &  \\
444   &  525  &  18:51:20.999  &  -06:14:01.33  &  0.249  $\pm$  0.001  &  15.93  &  14.48  &  0.43  &  0.0096  &  0.0762  &  b     &  \\
530   &  617  &  18:51:17.083  &  -06:13:39.61  &  0.95  $\pm$  0.01  &  16.19  &  13.48  &  1.97  &  0.0093  &  0.0980  &  n    &  \\
957   &  1075  &  18:51:04.289  &  -06:25:05.07  &  0.92  $\pm$  0.01  &  27.22  &  13.09  &  1.64  &  0.0090  &  0.0817  &  n    &  \\
1292  &  1433  &  18:50:55.568  &  -06:23:36.96  &  0.95  $\pm$  0.01  &  16.26  &  11.63  &  -0.08  &  0.0080  &  0.0444  &  a    &  \\
%1454  &  1600  &  18:50:48.813  &  -06:22:56.70  &  0.95  $\pm$  0.01  &  19.83  &  14.41  &  0.32  &  0.0078  &  0.0754  &  a    & y  \\
1509  &  1660  &  18:50:45.538  &  -06:12:54.69  &  2.19  $\pm$  0.08  &  22.77  &  14.57  &  0.54  &  0.0114  &  0.0512  &  a    &  \\
1606  &  1763  &  18:50:40.704  &  -06:14:44.25  &  1.23  $\pm$  0.02  &  16.58  &  14.03  &  0.66  &  0.0076  &  0.0383  &  n    &  \\
1647  &  1805  &  18:50:38.869  &  -06:13:12.26  &  0.94  $\pm$  0.01  &  19.10  &  12.92  &  1.59  &  0.0074  &  0.0742  &  n    &  \\
1663  &  1822  &  18:50:38.184  &  -06:13:51.35  &  1.18  $\pm$  0.02  &  16.42  &  12.20  &  1.05  &  0.0091  &  0.0647  &  n    &  \\
1719  &  1880  &  18:50:35.151  &  -06:12:50.15  &  5.4  $\pm$  0.4  &  17.45  &  13.76  &  0.52  &  0.0103  &  0.0319  &  n    &  \\
1733  &  1896  &  18:50:34.613  &  -06:25:05.34  &  20.0     $\pm$  0.7  &  40.27  &  14.28  &  2.00  &  0.0112  &  0.1112  &  n    &  \\
2098  &  ...    &  18:51:00.154  &  -06:14:49.72  &  0.434  $\pm$  0.009  &  28.45  &  12.37  &  0.00  &  0.0212  &  0.1073  &  b    &  \\
3662  &  5005  &  18:50:28.031  &  -06:10:09.14  &  0.95  $\pm$  0.01  &  23.17  &  14.20  &  1.74  &  0.0173  &  0.0596  &  n    &  \\
3860  &  5203  &  18:50:32.581  &  -06:14:57.60  &  5.5  $\pm$  0.4  &  22.06  &  14.80  &  1.74  &  0.0099  &  0.0729  &  n    &  \\
3867  &  5210  &  18:50:32.763  &  -06:20:04.30  &  1.05  $\pm$  0.01  &  32.74  &  13.95  &  1.98  &  0.0272  &  0.0918  &  n    &  \\
%3893  &  5236  &  18:50:33.439  &  -06:13:43.88  &  0.84  $\pm$  0.01  &  17.39  &  15.49  &  0.50  &  0.0300  &  0.0844  &  a    & y \\
3987  &  5330  &  18:50:35.453  &  -06:09:19.30  &  1.06  $\pm$  0.01  &  22.89  &  15.23  &  1.72  &  0.0197  &  0.0911  &  n    &  \\
4061  &  5404  &  18:50:37.144  &  -06:10:04.81  &  0.822   $\pm$  0.003  &  16.67  &  15.25  &  0.58  &  0.0167  &  0.0465  &  a    &  \\
4122  &  5465  &  18:50:38.537  &  -06:14:37.02  &  5.5  $\pm$  0.4  &  18.98  &  14.69  &  0.75  &  0.0111  &  0.0437  &  n    &  \\
4197  &  5540  &  18:50:40.578  &  -06:22:35.36  &  1.97  $\pm$  0.05  &  38.20  &  15.24  &  0.68  &  0.0202  &  0.0935  &  a    &  \\
4244  &  5587  &  18:50:41.695  &  -06:12:57.87  &  0.82  $\pm$  0.01  &  25.61  &  14.51  &  0.39  &  0.0157  &  0.0443  &  a    &  \\
4291  &  5634  &  18:50:42.840  &  -06:15:42.38  &  0.95  $\pm$  0.01  &  23.54  &  15.00  &  1.22  &  0.0141  &  0.1600  &  n    &  \\
4341  &  5684  &  18:50:43.663  &  -06:16:53.12  &  1.08  $\pm$  0.01  &  24.64  &  15.03  &  0.65  &  0.0122  &  0.0902  &  b    &  \\
4370  &  5713  &  18:50:44.146  &  -06:12:07.95  &  1.24  $\pm$  0.02  &  19.69  &  15.31  &  0.33  &  0.0362  &  0.0916  &  a    &  \\
4374  &  5717  &  18:50:44.235  &  -06:14:34.73  &  1.25  $\pm$  0.02  &  21.36  &  15.21  &  0.45  &  0.0346  &  0.0743  &  a    &  \\
4547  &  5890  &  18:50:47.746  &  -06:21:29.32  &  1.21  $\pm$  0.02  &  15.13  &  15.31  &  1.16  &  0.0132  &  0.0572  &  n    &  \\
4549  &  5892  &  18:50:47.769  &  -06:10:33.49  &  5.5  $\pm$  0.4  &  24.59  &  14.63  &  0.60  &  0.0201  &  0.0371  &  a    &  \\
4556  &  5899  &  18:50:47.853  &  -06:11:06.86  &  14 $\pm$  3  &  48.12  &  15.03  &  1.35  &  0.0159  &  0.2109  &  n    &  \\
5367  &  6710  &  18:51:03.950  &  -06:23:03.13  &  6.1  $\pm$  0.5  &  17.73  &  14.41  &  1.70  &  0.0074  &  0.0652  &  n    &  \\
5477  &  6820  &  18:51:06.144  &  -06:10:08.84  &  5.7  $\pm$  0.5  &  27.66  &  14.83  &  1.50  &  0.0131  &  0.1068  &  n    &  \\
6370  &  7713  &  18:51:22.449  &  -06:11:15.40  &  5.7  $\pm$  0.5  &  25.76  &  15.30  &  0.76  &  0.0173  &  0.0650  &  b    &  \\
6428  &  7771  &  18:51:23.829  &  -06:25:36.70  &  4.0  $\pm$  0.2  &  30.36  &  14.73  &  0.59  &  0.0099  &  0.0710  &  a    &  \\
6440  &  7783  &  18:51:24.012  &  -06:13:20.38  &  1.888   $\pm$  0.003  &  39.03  &  14.50  &  1.38  &  0.0090  &  0.0816  &  n    &  \\
6756  &  8099  &  18:51:31.434  &  -06:13:53.09  &  5.8  $\pm$  0.5  &  17.80  &  14.94  &  1.84  &  0.0154  &  0.0968  &  n    &  \\
8594   &  10095  &  18:51:10.248  &  -06:06:01.43  &  1.244 $\pm$  0.005  &  15.68  &  17.14  &  0.39  &  0.0609  &  0.1338  &  n    &  \\
9517   &  10696  &  18:51:41.247  &  -06:07:18.46  &  0.88  $\pm$  0.01  &  20.16  &  15.32  &  1.20  &  0.0129  &  0.1028  &  n    &  \\
10052  &  11078  &  18:51:19.119  &  -06:07:55.82  &  7.4  $\pm$  0.8  &  15.01  &  15.72  &  0.85  &  0.0197  &  0.0591  &  n    &  \\
10161  &  11155  &  18:50:41.777  &  -06:08:03.78  &  21.05 $\pm$  0.01  &  41.28  &  16.54  &  1.24  &  0.0552  &  0.3606  &  n    &  \\
10851  &  11629  &  18:50:55.996  &  -06:09:02.21  &  4.6  $\pm$  0.3  &  17.85  &  15.56  &  0.68  &  0.0146  &  0.0500  &  b    &  \\
10861  &  11637  &  18:50:54.708  &  -06:09:02.51  &  5.2  $\pm$  0.4  &  16.05  &  16.68  &  0.61  &  0.0568  &  0.1264  &  b    &  \\
%11595  &  12130  &  18:50:36.391  &  -06:09:58.80  &  0.87  $\pm$  0.01  &  24.73  &  15.96  &  0.63  &  0.0233  &  0.1176  &  b    & y \\
12485  &  12774  &  18:50:54.307  &  -06:11:09.85  &  5.2  $\pm$  0.4  &  30.33  &  15.79  &  0.60  &  0.0150  &  0.0691  &  b    &  \\
%12546  &  12816  &  18:51:41.118  &  -06:11:16.29  &  0.95  $\pm$  0.01  &  16.56  &  16.28  &  0.80  &  0.0359  &  0.0772  &  b    &  y\\
12646  &  12890  &  18:50:52.443  &  -06:11:23.12  &  2.6  $\pm$  0.1  &  23.89  &  15.74  &  0.55  &  0.0150  &  0.0575  &  b    &  \\
13073  &  13214  &  18:51:12.981  &  -06:11:58.35  &  1.22  $\pm$  0.02  &  15.74  &  15.78  &  1.46  &  0.0202  &  0.0809  &  n    &  \\
13511  &  13536  &  18:51:44.895  &  -06:12:34.27  &  0.327  $\pm$  0.001  &  33.78  &  16.28  &  0.46  &  0.0419  &  0.2369  &  b    &  \\
13540  &  13556  &  18:50:56.759  &  -06:12:35.98  &  5.3  $\pm$  0.4  &  15.95  &  15.57  &  0.81  &  0.0237  &  0.0752  &  n    &  \\
13951  &  13849  &  18:51:37.563  &  -06:13:11.24  &  1.25  $\pm$  0.02  &  18.47  &  16.95  &  0.79  &  0.0718  &  0.1666  &  b    &  \\
14095  &  13950  &  18:50:48.687  &  -06:13:24.10  &  3.5  $\pm$  0.1  &  18.25  &  16.18  &  0.57  &  0.0339  &  0.1054  &  b    &  \\
14450  &  14207  &  18:50:27.796  &  -06:13:55.76  &  2.5  $\pm$  0.1  &  24.80  &  16.53  &  0.90  &  0.0467  &  0.1442  &  b    &  \\
15291  &  14792  &  18:51:18.248  &  -06:15:18.63  &  0.83  $\pm$  0.01  &  15.54  &  16.05  &  1.13  &  0.0312  &  0.0781  &  n    &  \\
15633  &  15020  &  18:50:43.468  &  -06:15:50.88  &  5.6  $\pm$  0.4  &  18.04  &  15.66  &  1.58  &  0.0172  &  0.0428  &  n    &  \\
17415  &  16297  &  18:51:26.378  &  -06:18:44.88  &  0.420  $\pm$  0.002  &  18.54  &  16.03  &  0.59  &  0.0217  &  0.0801  &  b    &  \\
17546  &  16397  &  18:50:39.657  &  -06:18:53.31  &  0.83  $\pm$  0.01  &  16.12  &  17.00  &  1.10  &  0.0989  &  0.3246  &  n    &  \\
17649  &  16470  &  18:50:30.580  &  -06:19:03.98  &  0.362  $\pm$  0.002  &  22.86  &  16.89  &  0.94  &  0.2341  &  0.3098  &  b    &  \\
17730  &  16528  &  18:50:41.216  &  -06:19:05.59  &  5.2  $\pm$  0.4  &  22.95  &  16.67  &  1.12  &  0.0311  &  0.1987  &  n    &  \\
18776  &  17281  &  18:51:10.130  &  -06:20:29.10  &  4.8  $\pm$  0.3  &  18.16  &  16.52  &  0.61  &  0.0364  &  0.1463  &  b    &  \\
19357  &  17726  &  18:50:35.405  &  -06:21:13.05  &  1.18  $\pm$  0.02  &  17.35  &  16.25  &  0.69  &  0.0259  &  0.1019  &  b    &  \\
19570  &  17883  &  18:50:51.008  &  -06:21:31.20  &  0.83  $\pm$  0.01  &  22.48  &  15.87  &  0.52  &  0.0269  &  0.0753  &  b    &  \\
19844  &  18088  &  18:50:41.003  &  -06:21:53.06  &  0.2214  $\pm$  0.0007  &  40.20  &  15.98  &  1.16  &  0.0267  &  0.6817  &  n    & y \\
21243  &  19140  &  18:50:39.625  &  -06:23:35.99  &  0.488  $\pm$  0.003  &  15.20  &  16.15  &  0.70  &  0.1086  &  0.3077  &  b    &  \\
%22026  &  19708  &  18:50:50.242  &  -06:24:31.93  &  0.95  $\pm$  0.01  &  21.95  &  15.44  &  0.55  &  0.0201  &  0.0938  &  b    & y \\
22541  &  20108  &  18:50:53.732  &  -06:25:09.70  &  4.8  $\pm$  0.3  &  19.88  &  16.50  &  0.78  &  0.0346  &  0.1428  &  b    &  \\
23809  &  21055  &  18:51:17.696  &  -06:26:45.79  &  0.95  $\pm$  0.01  &  27.95  &  17.06  &  0.96  &  0.0380  &  0.1921  &  b    &  \\
23878  &  21104  &  18:50:36.522  &  -06:26:49.94  &  1.85  $\pm$  0.01  &  23.06  &  16.68  &  1.00  &  0.0495  &  0.3521  &  n    & y \\
25585  &  22164  &  18:50:30.755  &  -06:19:07.60  &  5.4  $\pm$  0.4  &  25.54  &  17.32  &  1.37  &  0.0697  &  0.2707  &  n    &  \\
26026  &  22451  &  18:50:33.299  &  -06:24:22.86  &  0.81  $\pm$  0.01  &  23.91  &  16.91  &  0.66  &  0.0561  &  0.1622  &  b    &  \\
26072  &  22483  &  18:50:29.833  &  -06:24:46.87  &  0.720  $\pm$  0.009  &  16.18  &  18.13  &  0.94  &  0.1793  &  0.3507  &  b    & \\
26158  &  22536  &  18:50:34.161  &  -06:25:35.99  &  3.2884  $\pm$  0.04  &  25.61  &  16.52  &  0.62  &  0.0839  &  0.4506  &  b    & y \\
27360  &  23360  &  18:51:01.356  &  -06:11:07.67  &  0.396  $\pm$  0.002  &  23.60  &  15.76  &  0.39  &  0.0296  &  0.1788  &  b    &  \\
29011  &  24388  &  18:51:45.297  &  -06:21:07.52  &  2.38  $\pm$  0.09  &  16.53  &  16.36  &  0.74  &  0.0744  &  0.1153  &  b    & y \\
29195  &  ...     &  18:50:27.943  &  -06:05:40.81  &  0.95  $\pm$  0.01  &  20.24  &  17.12  &  0.00  &  0.0918  &  0.2132  &  n    & y  \\
30166  &  25180  &  18:51:01.500  &  -06:18:41.95  &  1.23  $\pm$  0.02  &  20.62  &  15.47  &  0.40  &  0.0243  &  0.0715  &  b    &  \\
30525  &  25503  &  18:51:29.387  &  -06:15:22.32  &  0.639  $\pm$  0.006  &  32.82  &  16.41  &  0.82  &  0.0357  &  0.1778  &  b    &  \\
34773  &  26477  &  18:50:22.444  &  -06:13:50.70  &  1.23  $\pm$  0.4  &  18.01  &  16.20  &  0.72  &  0.0279  &  0.0886  &  b    &  \\
35526  &  26763  &  18:50:56.030  &  -06:09:42.59  &  5.0   $\pm$  0.3  &  27.36  &  16.73  &  1.14  &  0.0435  &  0.2130  &  n    &  \\
\hline
\enddata
\end{deluxetable}


\begin{thebibliography}{}
\bibitem[2001]{Bailer01} Bailer-Jones, C. A. L., \& Mundt, R. 2001, A\&A, 367, 218
\bibitem[2003]{Barnes03} Barnes, S.  2003, ApJ, 586, 464
\bibitem[2007]{Barnes07} Barnes, S.  2007, ApJ, 669, 1167
\bibitem[1997]{Bouvier97} Bouvier, J., Forestini, M., \& Allain, S.\ 1997, \aap, 326, 1023 
\bibitem[2000]{Cox00}Cox, A.N. 2000, in Allen's Astrophysical Quantities, 4th Edition, (Springer, AIP Press)
\bibitem[2000]{Girardi00} Girardi, L., Bressan, A., Bertelli, G., \& Chiosi, C. 2000, A\&AS, 141,371
\bibitem[2000]{Gonzalez00} Gonzalez, G., \& Wallerstein, G. 2000, PASP, 112, 1081
\bibitem[2003]{Guinan03} Guinan, E. F., McCook, G. P., DeWarf, L. E., et al. 2003, Bulletin of the American Astronomical Society, 35, 766
\bibitem[2006]{Dias06} Dias, W. S., Assafin, M., Fl\'orio, V., Alessi, B. S., \& Líbero, V. 2006, A\&A, 446, 949
%\bibitem[2008a]{Hartman08a} Hartman, J.D., Gaudi, B.S., Holman, M.J., et al. 2008a, ApJ, 675, 1233
%\bibitem[2008b]{Hartman08b} Hartman, J.D., Gaudi, B.S., Holman, M.J., et al. 2008b, ApJ, 675, 1254
\bibitem[2005]{Hargis05} Hargis, J.R., Sandquist, E.L., \& Bradsteert, D.H. 2005, AJ, 130, 2824
\bibitem[2009]{Hartman09} Hartman, J.D., Gaudi, B.S., Pinsonneault, M.H., et al. 2009, ApJ, 691, 342
\bibitem[1996]{Herbst96} Herbst, W. \& Wittenmyer, R. 1996, BAAS, 28, 1338
\bibitem[2002]{Herbst02} Herbst, W., Bailer-Jones, C. A. L., Mundt, R., Meisenheimer, K., \&  Wackermann, R., 2002, A\&A, 396, 513
\bibitem[2005]{Herbst05} Herbst, W., \& Mundt, R. 2005, ApJ, 633, 967
\bibitem[2006]{Hodgkin06} Hodgkin, S.T., Irwin, J.M., Aigrain, S., et al. 2006, AN, 327, 9
\bibitem[2007]{Holzwarth07} Holzwarth, V., \& Jardine, M.\ 2007, \aap, 463, 11 
\bibitem[1986]{Horne86} Horne, J. H. \& Baliunas, S. L. 1986, ApJ, 302, 757
\bibitem[2006]{Irwin06} Irwin, J., Aigrain, S., Hodgkin, S., et al. 2006, MNRAS, 370, 954
\bibitem[2007]{Irwin07} Irwin, J., Hodgkin, S., Aigrain, S., et al. 2007, MNRAS, 377, 741
\bibitem[2008]{Irwin08} Irwin, J., Hodgkin, S., Aigrain, S., et al. 2008, MNRAS, 383, 1588
\bibitem[2003]{Ivanova03} Ivanova, N., \& Taam, R.~E.\ 2003, \apj, 599, 516 
\bibitem[2007]{Kang07} Kang, Y.B., Kim, S.-L., Rey, S.-C., et al, 2007, PASP, 119, 239
\bibitem[1988]{Kawaler88} Kawaler, S.D., 1988, ApJ, 333, 236
%\bibitem[2001]{Kiss01} Kiss, L.L., Szab\`o, Gy.M., Szil\'adi, K., et al. 2001, A\&A, 376, 561
\bibitem[2001]{Kim01} Kim, S.-L., Chun, M.-Y., Park, B.-G. et al. 2001, A\&A 371,571
\bibitem[2007]{Koo07} Koo, J,-R., Kim, S.-L., Rey, S.-C. et al. 2007, PASP, 119, 1233
\bibitem[1981]{Kovacs81} Kovacs G., 1981, Ap\&SS, 78, 175 
\bibitem[1997]{Krishnamurthi97} Krishnamurthi, A., Pinsonneault, M.~H., Barnes, S., \& Sofia, S.\ 1997, \apj, 480, 303 
\bibitem[2004]{Lamm04} Lamm M.~H., Bailer-Jones C.~A.~L., Mundt R., Herbst W., Scholz A., 2004, A\&A, 417, 557 
\bibitem[1991]{MacGregor91} MacGregor, K.~B., \& Brenner, M.\ 1991, \apj, 376, 204 
\bibitem[2001]{Martin01}Martín, E.L., Dahm, S., \& Pavlenko, Y. 2001, ASP Conference Series Vol. 245. Ed. Ted von Hippel, Chris Simpson, and Nadine Manset. San Francisco, p.349
\bibitem[1986]{Mathieu86} Mathieu R.~D., \& Latham D.~W., 1986, AJ, 92, 1364 
\bibitem[1997]{Mcnamara77} McNamara, B.J., Pratt, N.M., \& Sanders, W.L. 1977, A\&AS, 27, 117
%\bibitem[1996]{Mermilliod96} Mermilliod, J.-C., Huestamendia, G., del Rio, G., \& Mayor, M.\ 1996, \aap, 307, 80
\bibitem[2009]{Meibom09} Meibom S., Mathieu R.~D., \& Stassun K.~G., 2009, ApJ, 695, 679
\bibitem[2001]{Messina01} Messina, S., Rodon\`o, M., \& Guinan, E. F. 2001, A\&A, 366, 215
\bibitem[2003]{Messina03} Messina, S., Pizzolato, N., Guinan, E. F., \& Rodon\'o, M. 2003, A\&A 410, 671
\bibitem[2007]{Messina07} Messina, S., 2007, Mem. Soc. Astron. It., 78, 628
\bibitem[2008a]{Messina08a} Messina, S., Distefano, E., Parihar, P., et al. 2008a,  A\&A, 483, 253
\bibitem[2008b]{Messina08b} Messina, S., 2008b,  A\&A, 480, 495
%\bibitem[]{} Messina, S. \& Guinan, E.F. 2002, A\&A, 393, 225
%\bibitem[]{} Messina, S. \& Guinan, E.F. 2003, A\&A, 409, 1017
\bibitem[2009]{Messina09} Messina, S., Desidera, S., Turatto, M., Lanzafame, A.C., \& Guinan, E.F. 2009, submitted to A\&A
\bibitem[1993]{Meynet93} Meynet,G., Mermilliod, J-.C., Maeder, A. 1993, A\&AS, 98, 477
%\bibitem[2002]{Nilakshi02} Nilakshi,, \& Sagar, R.\ 2002, \aap, 381, 65 
\bibitem[2009]{Parihar09} Parihar, P., Messina, S., Distefano, E.,  Shantikumar, N. S., \& Medhi, B. J. 2009,  MNRAS, 400, 603
%\bibitem[1998]{Perryman98} Perryman, M. A. C., Brown, A. G. A., Lebreton, Y., et al. 1998, A\&A, 331, 81
\bibitem[1992]{Press92} Press, W.H., Teukolsky, S.A., Vetterling, W.T., \& Flannery, B.P. 1992, in Numerical recipes in FORTRAN, Cambridge: University Press, 1992, 2nd ed.
\bibitem[1995]{Radick95}  Radick, R.R., Lockwood, G.W., Skiff, B.A., \& Thompson, D.T. 1995, ApJ, 452, 332
\bibitem[2004]{Rebull04} Rebull, L.M., Wolff, S.C., \& Strom, S.E. 2004, AJ, 127, 1029
\bibitem[1987]{Roberts87} Roberts, D. H., Lehar, J., \& Dreher, J. W. 1987, AJ, 93, 978
\bibitem[2007]{Roze07} Roze, M.B. \&  Hintz E.G., 2007, AJ, 134, 2067
%\bibitem[2000]{Rodono00} Rodon\`o, M., Messina, S., Lanza, A. F., Cutispoto, G., \& Teriaca, L. 2000, A\&A, 358, 624
\bibitem[1982]{Scargle82} Scargle, J.D. 1982, ApJ, 263, 835
\bibitem[2000]{Sills00} Sills, A., Pinsonneault, M.~H., \& Terndrup, D.~M.\ 2000, \apj, 534, 335 
\bibitem[1972]{Skumanich72} Skumanich, A. 1972, ApJ, 171, 565
\bibitem[1999]{Stassun99}  Stassun, K.G., Mathieu, R.D., Mazeh, T., \& Vrba, F. 1999, AJ, 117, 2941
\bibitem[1998]{Su98}Su C.-G., Zhao J.-L., \& Tian K.-P., 1998, A\&AS, 128, 255
\bibitem[1999]{Sung99} Sung,H., Bessel, M.S., Lee, H.-W., Kang, Y-H-. \& Lee, S.-W., 1999, MNRAS, 310, 982
\bibitem[2005]{von05} von Braun, K., Lee, B. L., Seager, S., et al. 2005, PASP, 117, 141 
\end{thebibliography}
\end{document}